\documentclass[english,a4paper,twocolumn,amsmath,amssymb,tightenlines,reprint]{revtex4-2}
\usepackage{bm}
\usepackage{graphicx,xcolor}
\definecolor{linkcolor}{HTML}{004191}
\usepackage[colorlinks,linkcolor=linkcolor,citecolor=linkcolor, urlcolor=linkcolor]{hyperref}
\usepackage{physics}
\usepackage{mlmodern}

\newcommand{\mean}[1]{\langle #1 \rangle}

\renewcommand{\vec}[1]{\bm #1}

\newcommand{\al}{\alpha}

\newcommand{\sig}{\sigma}

\newcommand{\x}{\vec r}

\newcommand{\im}{\text{i}}

\begin{document}

\title{Response to dynamic shape changes in suspensions of hard rectangles}

\author{Denis Dertli}
\author{Thomas Speck}
\email{thomas.speck@itp4.uni-stuttgart.de}
\affiliation{Institute for Theoretical Physics IV, University of Stuttgart, Heisenbergstr. 3, 70569 Stuttgart, Germany}

\begin{abstract}
	While the autonomous assembly of hard nanoparticles with different shapes has been studied extensively both in experiment and simulations, little is known about systems where particle shape can be dynamically altered. DNA origami nanostructures offer an alternative route to synthesize nanoparticles that can change their shape on demand. Motivated by recent experiments, here we study the structure and dynamics of suspensions of hard squares in response to an elongation into a rectangle. Performing extensive hard-particle Monte Carlo simulations at constant volume and employing two protocols, we numerically analyze the collective diffusion and ordering during quenching and the subsequent relaxation to the new equilibrium state. We find that the cascading protocol, which mimics experimentally realized DNA origami, can become dynamically arrested due to the increase in effective packing fraction.
\end{abstract}

\maketitle


\section{Introduction}

The interactions between hard particles are determined by their excluded volume and their shape. While in the simplest case these are spheres~\cite{royall24}, colloidal and nanoparticles with different shapes can now be synthesized routinely~\cite{sacanna13,bassani24}. In fact, through varying their shape an astonishingly wide range of ordered structures can be accessed in dense suspensions of hard particles~\cite{damasceno12}, which has been captured in theoretical approaches~\cite{keys11,wittkowski11} and rationalized through the emergence of directed entropic forces~\cite{vo22}. In particular in two dimensions, there is a rich transition behavior in hard-particle systems governed by shape and symmetry~\cite{anderson17,mizani25}. Hard-particle Monte Carlo (MC) simulations have been successfully utilized in order to study the equilibrium phase behavior of hard disks~\cite{bernard11,engel13,kapfer15}, $n$-gons~\cite{anderson17}, and hard rectangles~\cite{dertli25}.

However, typically colloidal particles are immutable after they have been synthesized with relatively few studies in which the particle shape has been changed dynamically~\cite{sun12,epstein15,tanjeem22}. While the response of dense suspensions to density and temperature quenches has been studied extensively, little is known about the response to a ``shape quench''.

DNA origami offers an alternative route to synthesize nanoparticles with almost arbitrary shapes~\cite{zhan23}. Essentially, a long single-stranded DNA scaffold is folded into a predetermined two- or three-dimensional nanostructure through hybridizing the scaffold with short complementary staple strands, which direct the folding and stabilize the final complex~\cite{rothemund06,dietz09,hong17}. Such DNA-based nanostructures are extensively researched motivated by their potential applications for therapy and drug delivery~\cite{endo13,sun22,he23,zhang23,safarkhani24,ding25}. Experimental studies on square and rod-like DNA origami nanostructures on lipid bilayers suggest that these can effectively behave as independent hard particles governed by Brownian motion~\cite{khmelinskaia21,fan25}.

\begin{figure*}[t]
	\includegraphics[scale=1]{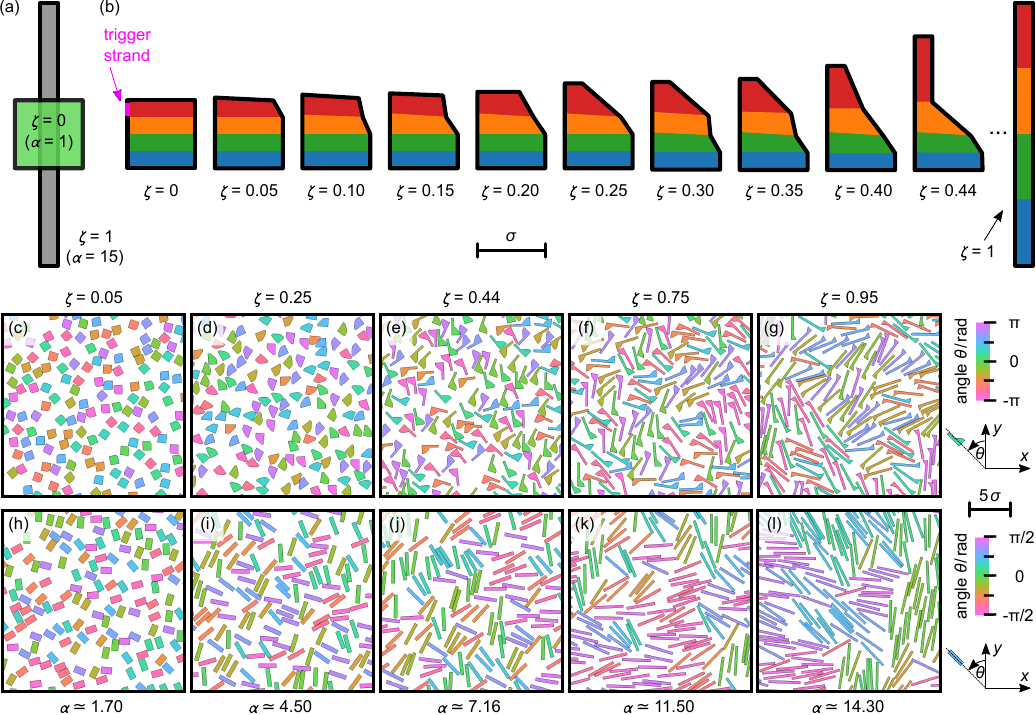}
	\caption{Shape transformation protocols: (a)~morphing and (b)~cascading. For the latter, the shape is divided into four sub-polygons with constant area each. The magenta arrow indicates the trigger strand in the upper left segment. (c--l)~Simulation snapshots representing different stages of a shape quench at packing fraction $\phi = 0.32$ for the (c--g)~cascading and (h--l)~morphing protocol. The colors indicate the orientation with respect to the $y$-axis.}
	\label{fig:models}
\end{figure*}

One of the key features of DNA origami is the opportunity to program different shapes and to dynamically reconfigure particle shape through the addition of trigger strands~\cite{zhang11,song17,wang20,kim23} that are designed to be more complementary to parts of the origami structure than the original staple strands. The trigger strands bind to short, exposed single-stranded ``toeholds'', initiating a cascade where they displace the original staples and thus cause a conformational change in the origami. In particular, a step-by-step relay process from a square to a rectangular shape has been implemented~\cite{song17,wang20}. This process has been exploited recently to achieve the reversible opening and closing of channels in synthetic cells due to collective forces exerted by the DNA nanostructures~\cite{fan25}. Inspired by these experiments, here we leverage computer simulations to study the collective behavior in suspensions of hard particles after triggering a change of their aspect ratio while keeping the area fixed. We compare two time-dependent protocols for the shape transformation and extend the established hard-particle MC simulation approach by incorporating these transformations: the cascading protocol following the experiments and a strongly simplified ``morphing'' protocol in which the initial squares are stretched over time [Fig.~\ref{fig:models}(a,b)]. Our numerical investigations focus on the diffusion and order properties in our many-particle systems. Both the dynamics during the shape change and the subsequent relaxation are addressed and compared to equilibrium.

\section{Simulation details}
\label{sec:sim}

We study a two-dimensional system of $N\simeq1300$ particles with fixed area $\sig^2$ at constant packing fraction $\phi$ and periodic boundary conditions~\cite{frenkel23}. A single configuration comprises the centroid positions $\x_k$ and the angles $\theta_k \in [-\pi/n , \pi/n)$ enclosed with the $y$-axis for each particle with $n$-fold rotational symmetry. Throughout, we employ dimensionless quantities with length unit $\sig$.

The particles interact through their excluded volume and we perform hard-particle MC simulations utilizing \texttt{HOOMD-blue}~\cite{anderson08,glaser15,anderson16,anderson20}. Only \emph{local} trial moves for translation and rotation are used that are either accepted or rejected if they lead to an overlap. Such dynamically realistic Monte Carlo schemes based on local moves are known to reproduce Brownian dynamics on intermediate and long time scales (see, e.g., Ref.~\cite{jack15}). We thus identify one MC time step as one translation or rotation (but not shape) trial move per particle with globally fixed and sufficiently small maximum size of translation and rotation for all runs (see Supplemental Information~\cite{sm} for further details).

In addition, we attempt shape changes. We consider two shape transformation protocols and introduce a shape parameter $\zeta \in [0,1]$ that tracks the actual shape: (i) the morphing protocol shown in Fig.~\ref{fig:models}(a), whereby the aspect ratio~$\alpha$ of a rectangle is successively increased from $\alpha = 1$ ($\zeta = 0$) to $\alpha = 15$ ($\zeta = 1$), i.e., $\zeta = (\alpha - 1)/14$ and (ii) the cascading protocol in Fig.~\ref{fig:models}(b) that also starts with a square shape ($\zeta = 0$) but captures some features of the toehold-mediated strand displacement process more accurately. During this protocol the outer particle shape is defined by a set of ten vertices that span four interior polygons, which is motivated by the cascading DNA origami process studied in Ref.~\cite{song17,wang20}. We suppose that the trigger strand corresponds to the left segment of the sub-polygon on the top. This segment starts to elongate, resulting in a deformation of the corresponding sub-polygon. Additionally, each sub-polygon is constrained to maintain a constant area $\sigma^2/4$. At $\zeta \simeq 0.44$, the sub-polygon on the top has reached its final shape. At the end of the cascading protocol ($\zeta=1$), particles have become rectangles with aspect ratio $\alpha=15$. For both protocols, a trial shape change is applied to all particles simultaneously and only accepted if no particle experiences an overlap so that the shape remains synchronous, i.e., all particles exhibit the same value $\zeta$.

Our simulations prepare equilibrated configurations at the target packing fraction. The shape transformation is triggered (defining $\tau = 0$) and we start collecting configurations. During this process, all particles either attempt local translation/rotation trial moves or a global small shape update advancing $\zeta$. The simulation is run until either the target shape ($\zeta = 1$) is reached or ``jamming'' occurs, i.e., further shape updates are accepted only very rarely (the acceptance ratio has dropped from initially one to below $10^{-6}$). We then start the relaxation run (defining $\tau' = 0$), where we collect further configurations without shape updates. If not stated otherwise, we calculate the mean~$\mean{\cdot}$ obtained from $\lesssim 30$ simulation runs with different random number seeds.


\section{Results and discussion}

\subsection{Quench of aspect ratio}

\begin{figure}[b!]
	\centering
	\includegraphics{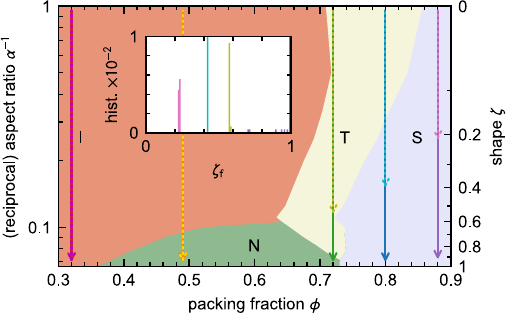}
	\caption{Quenches depicted in the phase diagram of hard rectangles. Five pairs of quenches from $\al=1$ (squares) to $\al=15$ (rectangles) are investigated at packing fractions $\phi = 0.32$, $0.49$, $0.72$, $0.80$ and $0.88$. The solid and dotted arrows correspond to the morphing and cascading protocol, respectively. The underlying phase diagram is taken from Ref.~\cite{dertli25} comprising isotropic fluid (I), nematic (N), smectic (S), and tetratic (T) order. For the morphing protocol there is a direct correspondence from $\zeta$ (right axis) to aspect ratio $\al$. The inset shows the distribution of the final shape parameter values~$\zeta_\text{f}$ obtained from each simulation run. The target shape is reached ($\mean{\zeta_\text{f}} = 1$) except for: morphing protocol with $\phi = 0.88$ (purple) and cascading protocol with $\phi = 0.72$ (olive), $0.80$ (cyan) and $0.88$ (pink). Arrow heads terminate at $\mean{\zeta_\text{f}}$.}
	\label{fig:phasediagQuenches}
\end{figure}

\begin{figure}[t]
	\centering
	\includegraphics{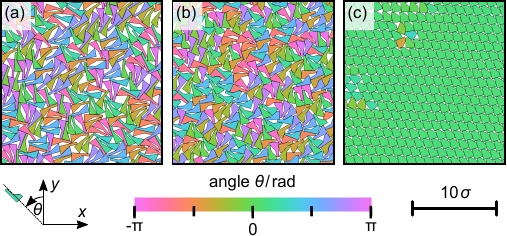}
	\caption{Simulation snapshots showing the jamming ($\zeta \simeq \zeta_\text{f} \simeq \text{const.} < 1$, $\tau \gtrsim 10^7$) observed at the termination of the high packing fraction quenches within the cascading protocol: (a) $\phi = 0.72$ ($\zeta_\text{f} \simeq 0.578$), (b) $\phi = 0.80$ ($\zeta_\text{f} \simeq 0.423$), and (c) $\phi = 0.88$ ($\zeta_\text{f} \simeq 0.233$).}
	\label{fig:jamming}
\end{figure}

For the quenches at low packing fraction ($\phi = 0.32$), the collective behavior within both protocols is shown in the sequences of simulation snapshots in Fig.~\ref{fig:models}. A crucial difference between the morphing and cascading protocols is that only during the morphing rectangles always exhibit a two-fold rotational symmetry and thus are in principle space-filling. In total, we study five pairs of quenches at different packing fractions shown within the phase diagram of hard rectangles in Fig.~\ref{fig:phasediagQuenches} that we obtained previously~\cite{dertli25}. In equilibrium, hard rectangles exhibit four phases: isotropic fluid (I), nematic (N), smectic (S), and tetratic (T) order. While for the low packing fraction quenches ($\phi \leq 0.49$) the shape transformations always succeed (i.e., $\zeta = 1$ reaches unity and particles have adopted the target shape), in the high packing fraction regime ($\phi > 0.49$) we observe $\zeta = \zeta_\text{f} < 1$. The distribution of the final shape parameter values $\zeta_\text{f}$ is very narrow as shown in the inset in Fig.~\ref{fig:phasediagQuenches}. In Fig.~\ref{fig:jamming}, we show representative snapshots of configurations taken at the point when the final shape has been reached for the packing fraction where the final shape is not the target shape ($\zeta_\text{f} < 1$). For the morphing protocol at the highest packing fraction $\phi = 0.88$, a 	finite-size analysis~\cite{sm} indicates that eventually $\zeta_\text{f} \to 1$ is reached. Interestingly, the sharp termination of the cascading protocol at values smaller than unity seem not to be subject to finite-size effects and occur at the same values of $\zeta_\text{f}$ independent of system size.

\subsection{Mean-square displacement}

Our aim is to understand how a shape transformation impacts the average dynamics of particles. As reference, we first quantify the equilibrium diffusion coefficient $D_\text{eff}$ obtained from the linear growth of the mean-squared displacement (MSD) with respect to MC time $\tau$ through
\begin{equation}
	r^2 (\tau) = 2 d D_\text{eff} \tau
	\label{eq:msdModel}
\end{equation}
in $d=2$ dimensions. Given the trajectories $\{ \vec r_k(\tau) \}_{k=1}^{N}$, we can calculate this quantity by exploiting both particle and time interval averages
\begin{equation}
	r^2 (\tau) = \left\langle \frac{1}{N}\sum_{k=1}^N 
	\frac{1}{I_{\tau}}\sum_{i=0}^{I_{\tau}-1} \left[ \vec 
	r_k(t_i + \tau) - \vec r_k(t_i) \right]^2 \right\rangle
	\label{eq:msdDefEq}
\end{equation}
with the number $I_{\tau} = (\tau_\mathrm{f} - \tau) / \Delta\tau + 1$ of time intervals of length~$\tau$, the sampled time points $t_i = i \times \Delta \tau$, the observation step size $\Delta \tau \simeq 10^4$, and the final MC time point $\tau_\mathrm{f} \simeq 2 \times 10^8$. For calculating MSDs, we employ the package \texttt{freud}~\cite{ramasubramani20}.

\begin{figure}[t]
	\centering
	\includegraphics{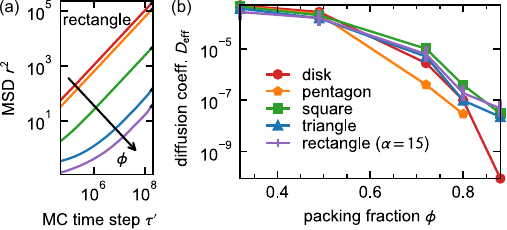}
	\caption{Equilibrium diffusion behavior as function of packing fraction and shape. (a)~MSD based on Eq.~\eqref{eq:msdDefEq} for hard rectangles with aspect ratio $\alpha=15$ at packing fractions $\phi = 0.32$ (red), $0.49$ (orange), $0.72$ (green), $0.80$ (blue), and $0.88$ (purple). (b)~Long-time effective diffusion coefficient~$D_\text{eff}$ for hard disks (red), pentagons (orange), squares (green), triangles (blue), and rectangles with aspect ratio $\alpha=15$ (purple). $D_\text{eff}$ has been estimated fitting the linear regime in (a) according to Eq.~\eqref{eq:msdModel} for the hard rectangles (and analogously for the other shapes~\cite{sm}). For pentagons, compression to the highest target packing fraction ($\phi = 0.88$) was not reached within finite compression times (of the same order of magnitude as for the other shapes).}
	\label{fig:eqmsd}
\end{figure}

\begin{figure}[t]
	\centering
	\includegraphics{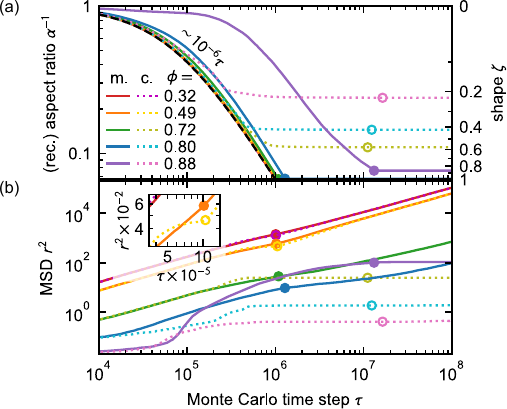}
	\caption{Evolution of (a)~the shape parameter~$\zeta$ and (b)~the MSD [Eq.~\eqref{eq:msdDefNotEq}] for several packing fractions $\phi$ and both the morphing (m) and cascading (c) protocols. The disks and circles indicate the time when the protocol reaches the final shape and no shape updates are attempted beyond this time. The black dashed line indicates the linear regime $\zeta \sim \tau$ that is obeyed by the low packing fraction quenches. The inset displays a zoom for packing fraction $\phi = 0.49$ when reaching the final shape. Colors and line styles are consistent with Fig.~\ref{fig:phasediagQuenches}.}
	\label{fig:zetaop24msd}
\end{figure}

Intuitively, we expect that higher packing fractions~$\phi$ are associated with a slower growth of the MSD. This trend is confirmed by the motion of hard rectangles in Fig.~\ref{fig:eqmsd}(a). Another crucial impact on the available area to move is the shape of the particles as it defines the effectively excluded area. The diffusion coefficients $D_\text{eff}$ for different shapes (with the same area) are compared in Fig.~\ref{fig:eqmsd}(b). While the packing fraction is the dominant factor (approximately setting the order of magnitude), the shape leads to a ``fine splitting'' of the $D_\text{eff}$ spectrum. For the lowest packing fraction ($\phi = 0.32$), disks exhibits the highest value of $D_\text{eff}$ in comparison to the other shapes. A closer inspection reveals the tendency that the larger the deviation from a disk shape (in terms of both a larger circumference and higher anisotropy with respect to rotational symmetry), the slower the dynamics becomes with the rod-like particle (rectangle of aspect ratio~$\alpha=15$) being the slowest one. Intriguingly, at higher packing fractions this ordering of shape changes, and we find that at the highest packing fraction ($\phi = 0.88$) disks are several orders of magnitudes slower than the rectangle. We attribute this change to the fact that shapes with an axial symmetry exhibit a further degree of freedom, allowing them to collectively align and reduce the excluded area cooperatively. Another fundamental aspect that comes into play is the non-space-filling geometrical character of various shapes (e.g., disks), i.e., there is a maximal packing fraction above which an overlap-free configuration cannot exist anymore.

During the shape-changing protocols introduced in Sec.~\ref{sec:sim} and the subsequent relaxation, the MSD not only depends on the relative time different but also the absolute time. For that reason, exploiting time averages as in Eq.~\eqref{eq:msdDefEq} is not feasible, and we have to adopt the definition of the MSD to
\begin{equation}
	r^2 (\tau) = \left\langle \frac{1}{N}\sum_{k=1}^N \left[ \vec r_k(\tau) - 
	\vec r_k(0) \right]^2 \right\rangle,
	\label{eq:msdDefNotEq}
\end{equation}
which can be formally understood as the special case of Eq.~\eqref{eq:msdDefEq} with a single time interval by interpreting $\tau = \tau_\mathrm{f}$. Figure~\ref{fig:zetaop24msd}(a,b) shows the evolution of the shape parameter $\zeta$ and the MSD, respectively, as a function of MC time during the five pairs of quenches studied. In the low packing fraction regime ($\phi \leq 0.49$), both protocols are in close accordance and show normal diffusion. The morphing protocol at the highest packing fraction $\phi = 0.88$ exhibits a slow evolution of shape but a faster increase of the MSD than at $\phi = 0.80$, which can be attributed to the higher diffusion coefficient for squares. The inset in Fig.~\ref{fig:zetaop24msd}(b) zooms onto the quenches for $\phi = 0.49$, where we see a pronounced slow-down of the cascading dynamics before the target shape ($\zeta = 1$) is reached eventually. We find that $\phi = 0.49$ is the highest packing fraction for which a full completion of the transformation ($\zeta = 1$) is achieved. At higher packing fractions ($\phi > 0.49$), the cascading protocol does not recover from the slow-down and the dynamics is arrested, resulting in a flattening of shape [Fig.~\ref{fig:zetaop24msd}(a)] and MSD [Fig.~\ref{fig:zetaop24msd}(b)] curves reaching a constant value.

\begin{figure}[t]
	\centering
	\includegraphics{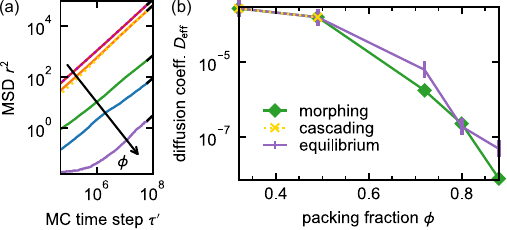}
	\caption{Diffusion behavior after quenching. (a)~MSD based on Eq.~\eqref{eq:msdDefNotEq} for constant shape ($\zeta = \text{const.}$) following a quench. Colors and line styles are consistent with Fig.~\ref{fig:phasediagQuenches}. (b)~Long-time effective diffusion coefficient $D_\text{eff}$ for hard rectangles (with aspect ratio $\alpha \simeq 15$) originating from the morphing protocol (green) and shapes stemming from the cascading protocol (gray). In the latter protocol, hard rectangles (with aspect ratio $\alpha \simeq 15$) are only obtained for low packing fractions $\phi \leq 0.49$ (cf. inset in Fig.~\ref{fig:phasediagQuenches}). As reference, we show the equilibrium values of $D_\text{eff}$ for hard rectangles with aspect ratio $\alpha = 15$ (purple) that have been initialized at a high packing fraction in the smectic phase without undergoing a transformation process (cf. Fig.~\ref{fig:eqmsd}).}
	\label{fig:eqVsOrigami}
\end{figure}

For the morphing protocol, we also observe a flattening of the MSD for the highest packing fraction $\phi = 0.88$ [Fig.~\ref{fig:zetaop24msd}(b)], apparently suggesting that the dynamics become arrested. To investigate the dynamics after quenching, we perform additional simulations without shape changes, for which we take the final quench configurations as initial configurations. We calculate the MSD with respect to this reference frame ($\tau'=0$) and recover normal diffusion in the long-time limit [Fig.~\ref{fig:eqVsOrigami}(a)]. For low packing fractions ($\phi \leq 0.49$) are the MSDs of both protocols in Fig.~\ref{fig:eqVsOrigami}(a) in close accordance, cf. Fig.~\ref{fig:eqmsd}. For $\phi > 0.49$ the MSD is essentially flat for the cascading protocol and the diffusion coefficient vanishes also in the continued simulations without shape changes. We conclude that for high packing fractions ($\phi > 0.49$), the simulation path taken plays a fundamental role and -- even for relatively long relaxation times $\Delta \tau \sim 10^8$ -- the ``memory'' of the system does not forget its path. For instance, the diffusion coefficient $D_\text{eff}$ for hard rectangles (aspect ratio $\alpha = 15$) clearly differs for $\phi > 0.49$. This observation can be understood by taking into account that the changing the shape can result in a metastable state that exhibits another phase than the actual equilibrium phase. Likewise, at $\phi = 0.72$, rod-like rectangles ($\alpha = 15$) self-assemble in a (uniaxial) nematic phase in equilibrium, whereas after morphing the system is in a metastable state, featuring a biaxial order. The higher value of $D_\text{eff}$ in the former case is reasonable, because the nematic phase enhances parallel diffusion (along the director), where the rod-like particles can slide past each other. Obviously, this kind of movement is prevented in the biaxial order of the metastable state. Intriguingly, $D_\text{eff}$ is in close accordance at $\phi = 0.80$ for rectangles ($\alpha = 15$), although the order depending on the path differs: We recapitulate that after morphing the system does not result in a smectic phase (as naively expected from Fig.~\ref{fig:phasediagQuenches}), but rather remains in a biaxial metastable state.

\subsection{Evolution of orientational order}

\begin{table}[b!]
	\caption{Idealized values of the $\nu$-fold orientational order parameters~$\psi_{\nu}$ [Eq.~\eqref{eq:psi:k}] characterizing the isotropic~(I), tetratic~(T), nematic~(N) or smectic~(S) phase. The actual equilibrium values for hard rectangles with various aspect ratios~$\alpha \in (1,15]$ are summarized in the Supplemental Information~\cite{sm}.}
	\label{tab:psi:k}
	\begin{center}
		\begin{tabular}{l l l}
			\hline\hline
			phase \hspace{1.9cm} &  $\psi_4$ \hspace{1.9cm} &  $\psi_2$  \\
			\hline
			I  & \hspace{0.015cm} $0$ & \hspace{0.015cm} $0$  \\
			T  & \hspace{0.015cm} $1$ & \hspace{0.015cm} $0$  \\
			N/S  & \hspace{0.015cm} $1$ & \hspace{0.015cm} $1$  \\
			\hline\hline
		\end{tabular}
	\end{center}
\end{table}

\begin{figure}[t]
	\centering
	\includegraphics{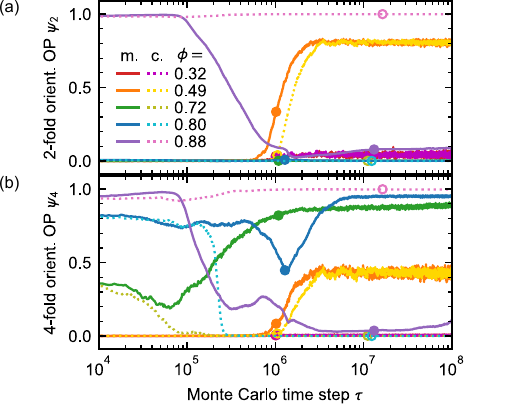}
	\caption{Evolution of the orientational order parameters (a)~$\psi_2$ and (b)~$\psi_4$ defined in Eq.~\eqref{eq:psi:k} following the evolution of the shape shown in Fig.~\ref{fig:zetaop24msd}(a).}
	\label{fig:op24}
\end{figure}

To gain further insights into the ordering of particles during the shape quench and the subsequent relaxation, we employ the two order parameters
\begin{equation}
	\psi_\nu = \left\langle \left|\frac{1}{N}\sum_{k=1}^N 
	e^{\im\nu\theta_k}\right|^2 \right\rangle
	\label{eq:psi:k}
\end{equation}
with $\nu=2,4$ and $0\leqslant\psi_\nu\leqslant1$ measuring orientational order. Their logical conjunction allows us to distinguish the different phases according to Tab.~\ref{tab:psi:k}. The temporal evolution of the order parameters is shown in Fig.~\ref{fig:op24}(a,b). Initially, for each pair of quenches the order coincides as both protocols start from the same pre-relaxed system of squares. In accordance with our conclusions for the diffusion behavior, we find that the orientational order during both protocols qualitatively agrees for the low packing fraction quenches ($\phi \leq 0.49$), where the systems remain isotropic ($\psi_2 \simeq \psi_4 \simeq 0$) as in equilibrium (cf. Fig.~\ref{fig:phasediagQuenches} and Supplemental Information~\cite{sm}). Notably, the expected final nematic phase is absent during the cascading quench and only manifests once the caging has vanished, whereas the emergence of this phase is already observed during the finalization of the morphing quench. This observation is qualitatively rationalized by the lower degree of rotational symmetry of the cascading shapes. All low packing fraction quenches ($\phi \leq 0.49$) exhibit the same orientational order during relaxation and reach the stable equilibrium state~\cite{sm}.

\begin{figure}[t]
	\centering
	\includegraphics{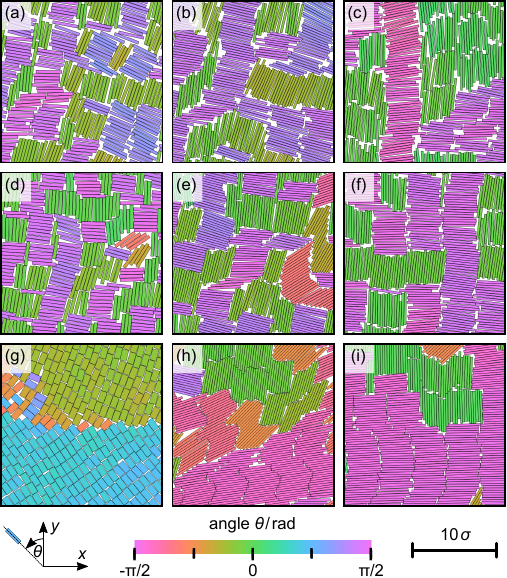}
	\caption{Sequences of simulation snapshots showing the temporal evolution during the morphing protocol for the high packing fraction quenches: (a)-(c) $\phi = 0.72$ ($\tau \simeq 6.5 \times10^5$, $1.1 \times10^6$, $1.0 \times10^8$; $\zeta \simeq 0.59$, $0.99$, $1.00$), (d)-(f) $\phi = 0.80$ ($\tau \simeq 6.5 \times10^5$, $1.3 \times10^6$, $1.0 \times10^8$; $\zeta \simeq 0.48$, $1.00$, $1.00$), and (g)-(i) $\phi = 0.88$ ($\tau \simeq 6.5 \times10^5$, $9.3 \times10^6$, $1.1 \times10^8$; $\zeta \simeq 0.05$, $1.00$, $1.00$).}
	\label{fig:simpleHigh}
\end{figure}

At packing fraction $\phi = 0.72$, quenches start in the tetratic phase and their $\psi_4$ signals diverge early ($\tau \sim 5 \times 10^4$). While there is still normal diffusion, the emerging non-rectangular shapes of the cascading protocol show an early replacement of the tetratic phase by the isotropic phase [cf. Fig.~\ref{fig:jamming}(a)]. The observed behavior during the morphing protocol is different, where $\psi_4$ essentially scans the equilibrium state point $(\alpha,\phi=0.72)$ values of rectangles with aspect ratio $\alpha$~\cite{sm}. In contrast to equilibrium, $\psi_2$ does not show any signal, i.e., the expected final nematic order is absent and not reached even after long relaxation. Accordingly, the TN transition is suppressed, which can be interpreted as a further indicator for its discontinuous character as suggested in Ref.~\cite{dertli25}. The sequence of snapshots in Fig.~\ref{fig:simpleHigh}(a)-(c) illustrates the evolution into this final metastable state that exhibits a biaxial orientation order.

Qualitatively, increasing the packing fraction to $\phi = 0.80$ quenches behave similar to the previous case except for an effect occurring at the end of the morphing quench: In contrast to the equilibrium expectation ($\psi_4 \simeq 1$~\cite{sm}) the value of $\psi_4$ drops. However, it rises again and saturates towards its equilibrium value as soon as the shape transformation is accomplished. Interestingly, this temporal drop of $\psi_4$ occurs in the neighborhood of the equilibrium TS transition parameter range [cf. Fig.~\ref{fig:phasediagQuenches}, \ref{fig:zetaop24msd}(a,d)], which is also suppressed ($\psi_2 \simeq 0$) in favor of a metastable tetratic phase -- in accordance with the strong hysteresis for the equilibrium TS transition discussed in Ref.~\cite{dertli25}. A visual inspection of Fig.~\ref{fig:simpleHigh}(d)-(f) reveals that the temporal drop of $\psi_4$ can be attributed to the emergence of defects, which exhibit the character of smectic domains that align in the further progress according to the two favored directions of the overall biaxial order. We note that our finite-size analysis points to the tendency that this drop of $\psi_4$ becomes slightly weaker as the system size increases~\cite{sm}. 

Finally, starting from even more densely packed squares at $\phi = 0.88$, the cascading protocol jams early and remains uniaxial [$\psi_2 \simeq \psi_4 \simeq 1$, cf. Figs.~\ref{fig:jamming}(c) and \ref{fig:zetaop24msd}(c,d)]. On the contrary, the morphing protocol succeeds in the transformation but shows a non-monotonic behavior at intermediate times before terminating at $\psi_2 \simeq \psi_4 \simeq 0$, which is the opposite of the equilibrium expectation ($\psi_2 \simeq \psi_4 \simeq 1$~\cite{sm}). Thus, in the morphing $\phi = 0.88$ quench neither a global uni- nor biaxial phase manifests, also not after a long relaxation. Figure~\ref{fig:simpleHigh}(g-i) suggest that interfaces are formed due to the non-commensurability (i.e., an incompatible simulation box aspect ratio) alongside defects that lead to the emergence of competing local smectic clusters without a distinct global orientational order. It is conceivable that the defects serve as nucleation sites, which are frozen into a kinetically trapped state.

\section{Conclusions}

To conclude, we have numerically investigated the response of hard rectangles to dynamic shape changes. We compare two protocols for the same initial and final aspect ratio: (i) the simplified morphing protocol, whereby the aspect ratio of the rectangles is successively increased, and (ii) the cascading protocol mimicking the experimental process more closely. For the cascading protocol, we find a transition between two regimes: For low packing fractions ($\phi \leq 0.49$) the dynamics and ordering during the shape quench and in the stationary regime are in close agreement and recover equilibrium properties. For high packing fractions ($\phi > 0.49$) the evolution of the shape cannot proceed to the target shape and the system becomes dynamically arrested as evident in the mean-square displacement, which levels off to a plateau. Due to the absence of this arrest during the morphing protocol (going through a sequence of space filling rectangles with increasing aspect ratio) even for the highest packing fraction ($\phi = 0.88$), we attribute the observed jamming of particles to the non-space-filling nature of the cascading shapes effectively decreasing the available area and thus increasing the effective packing fraction. For the morphing protocol the tetratic phase becomes predominant both during the evolution and in the stationary regime: In particular, a shape quench starting in the tetratic phase and terminating within the phase boundaries of the nematic phase ($\phi = 0.72$) remains tetratic even after a long relaxation of $\Delta \tau \simeq 10^8$ MC time steps. Moreover, we find that a shape quench ($\phi = 0.80$) that crosses the tetratic-smectic transition in the phase diagram of hard rectangles does not result in the smectic phase, but rather in a metastable tetratic phase and supports the identification of the tetratic-smectic transition as discontinuous in Ref.~\cite{dertli25}.

In the quest for functional and adaptive soft materials and their autonomous assembly, designing programmable and dynamically changing interactions has moved into the focus, e.g. through exploiting allosteric transitions~\cite{lazaro16,metson25}. For (almost) hard particles, adaptive interactions can be achieved through adjusting the particle shape~\cite{tanjeem22}. Our numerical study is a first step towards understanding how such shape changes might be exploited to design and access long-lived metastable collective states.


\section*{Conflicts of interest}

There are no conflicts of interest to declare.


\begin{acknowledgments}
	Simulations have been performed using the bwHPC infrastructure (clusters BinAC and BinAC2) supported by the state of Baden-Württemberg and the Deutsche Forschungsgemeinschaft (DFG) through grant INST 37/935-1 and 37/1159-1 FUGG. In addition, we gratefully acknowledge computing time granted by SimTech (University of Stuttgart and DFG grant no. 390740016 -- EXC 2075).
\end{acknowledgments}


%


\begin{thebibliography}{42}%
\makeatletter
\providecommand \@ifxundefined [1]{%
 \@ifx{#1\undefined}
}%
\providecommand \@ifnum [1]{%
 \ifnum #1\expandafter \@firstoftwo
 \else \expandafter \@secondoftwo
 \fi
}%
\providecommand \@ifx [1]{%
 \ifx #1\expandafter \@firstoftwo
 \else \expandafter \@secondoftwo
 \fi
}%
\providecommand \natexlab [1]{#1}%
\providecommand \enquote  [1]{``#1''}%
\providecommand \bibnamefont  [1]{#1}%
\providecommand \bibfnamefont [1]{#1}%
\providecommand \citenamefont [1]{#1}%
\providecommand \href@noop [0]{\@secondoftwo}%
\providecommand \href [0]{\begingroup \@sanitize@url \@href}%
\providecommand \@href[1]{\@@startlink{#1}\@@href}%
\providecommand \@@href[1]{\endgroup#1\@@endlink}%
\providecommand \@sanitize@url [0]{\catcode `\\12\catcode `\$12\catcode `\&12\catcode `\#12\catcode `\^12\catcode `\_12\catcode `\%12\relax}%
\providecommand \@@startlink[1]{}%
\providecommand \@@endlink[0]{}%
\providecommand \url  [0]{\begingroup\@sanitize@url \@url }%
\providecommand \@url [1]{\endgroup\@href {#1}{\urlprefix }}%
\providecommand \urlprefix  [0]{URL }%
\providecommand \Eprint [0]{\href }%
\providecommand \doibase [0]{https://doi.org/}%
\providecommand \selectlanguage [0]{\@gobble}%
\providecommand \bibinfo  [0]{\@secondoftwo}%
\providecommand \bibfield  [0]{\@secondoftwo}%
\providecommand \translation [1]{[#1]}%
\providecommand \BibitemOpen [0]{}%
\providecommand \bibitemStop [0]{}%
\providecommand \bibitemNoStop [0]{.\EOS\space}%
\providecommand \EOS [0]{\spacefactor3000\relax}%
\providecommand \BibitemShut  [1]{\csname bibitem#1\endcsname}%
\let\auto@bib@innerbib\@empty
\bibitem [{\citenamefont {Royall}\ \emph {et~al.}(2024)\citenamefont {Royall}, \citenamefont {Charbonneau}, \citenamefont {Dijkstra}, \citenamefont {Russo}, \citenamefont {Smallenburg}, \citenamefont {Speck},\ and\ \citenamefont {Valeriani}}]{royall24}%
  \BibitemOpen
  \bibfield  {author} {\bibinfo {author} {\bibfnamefont {C.~P.}\ \bibnamefont {Royall}}, \bibinfo {author} {\bibfnamefont {P.}~\bibnamefont {Charbonneau}}, \bibinfo {author} {\bibfnamefont {M.}~\bibnamefont {Dijkstra}}, \bibinfo {author} {\bibfnamefont {J.}~\bibnamefont {Russo}}, \bibinfo {author} {\bibfnamefont {F.}~\bibnamefont {Smallenburg}}, \bibinfo {author} {\bibfnamefont {T.}~\bibnamefont {Speck}},\ and\ \bibinfo {author} {\bibfnamefont {C.}~\bibnamefont {Valeriani}},\ }\bibfield  {title} {\bibinfo {title} {Colloidal hard spheres: {{Triumphs}}, challenges, and mysteries},\ }\href {https://doi.org/10.1103/RevModPhys.96.045003} {\bibfield  {journal} {\bibinfo  {journal} {Rev. Mod. Phys.}\ }\textbf {\bibinfo {volume} {96}},\ \bibinfo {pages} {045003} (\bibinfo {year} {2024})}\BibitemShut {NoStop}%
\bibitem [{\citenamefont {Sacanna}\ \emph {et~al.}(2013)\citenamefont {Sacanna}, \citenamefont {Korpics}, \citenamefont {Rodriguez}, \citenamefont {{Col{\'o}n-Mel{\'e}ndez}}, \citenamefont {Kim}, \citenamefont {Pine},\ and\ \citenamefont {Yi}}]{sacanna13}%
  \BibitemOpen
  \bibfield  {author} {\bibinfo {author} {\bibfnamefont {S.}~\bibnamefont {Sacanna}}, \bibinfo {author} {\bibfnamefont {M.}~\bibnamefont {Korpics}}, \bibinfo {author} {\bibfnamefont {K.}~\bibnamefont {Rodriguez}}, \bibinfo {author} {\bibfnamefont {L.}~\bibnamefont {{Col{\'o}n-Mel{\'e}ndez}}}, \bibinfo {author} {\bibfnamefont {S.-H.}\ \bibnamefont {Kim}}, \bibinfo {author} {\bibfnamefont {D.~J.}\ \bibnamefont {Pine}},\ and\ \bibinfo {author} {\bibfnamefont {G.-R.}\ \bibnamefont {Yi}},\ }\bibfield  {title} {\bibinfo {title} {Shaping colloids for self-assembly},\ }\href {https://doi.org/10.1038/ncomms2694} {\bibfield  {journal} {\bibinfo  {journal} {Nat. Commun.}\ }\textbf {\bibinfo {volume} {4}},\ \bibinfo {pages} {1688} (\bibinfo {year} {2013})}\BibitemShut {NoStop}%
\bibitem [{\citenamefont {Bassani}\ \emph {et~al.}(2024)\citenamefont {Bassani}, \citenamefont {Van~Anders}, \citenamefont {Banin}, \citenamefont {Baranov}, \citenamefont {Chen}, \citenamefont {Dijkstra}, \citenamefont {Dimitriyev}, \citenamefont {Efrati}, \citenamefont {Faraudo}, \citenamefont {Gang}, \citenamefont {Gaston}, \citenamefont {Golestanian}, \citenamefont {{Guerrero-Garcia}}, \citenamefont {Gruenwald}, \citenamefont {{Haji-Akbari}}, \citenamefont {Ib{\'a}{\~n}ez}, \citenamefont {Karg}, \citenamefont {Kraus}, \citenamefont {Lee}, \citenamefont {Van~Lehn}, \citenamefont {Macfarlane}, \citenamefont {Mognetti}, \citenamefont {Nikoubashman}, \citenamefont {Osat}, \citenamefont {Prezhdo}, \citenamefont {Rotskoff}, \citenamefont {Saiz}, \citenamefont {Shi}, \citenamefont {Skrabalak}, \citenamefont {Smalyukh}, \citenamefont {Tagliazucchi}, \citenamefont {Talapin}, \citenamefont {Tkachenko}, \citenamefont {Tretiak}, \citenamefont {Vaknin}, \citenamefont {{Widmer-Cooper}}, \citenamefont {Wong}, \citenamefont {Ye}, \citenamefont {Zhou}, \citenamefont {Rabani}, \citenamefont {Engel},\ and\ \citenamefont {Travesset}}]{bassani24}%
  \BibitemOpen
  \bibfield  {author} {\bibinfo {author} {\bibfnamefont {C.~L.}\ \bibnamefont {Bassani}}, \bibinfo {author} {\bibfnamefont {G.}~\bibnamefont {Van~Anders}}, \bibinfo {author} {\bibfnamefont {U.}~\bibnamefont {Banin}}, \bibinfo {author} {\bibfnamefont {D.}~\bibnamefont {Baranov}}, \bibinfo {author} {\bibfnamefont {Q.}~\bibnamefont {Chen}}, \bibinfo {author} {\bibfnamefont {M.}~\bibnamefont {Dijkstra}}, \bibinfo {author} {\bibfnamefont {M.~S.}\ \bibnamefont {Dimitriyev}}, \bibinfo {author} {\bibfnamefont {E.}~\bibnamefont {Efrati}}, \bibinfo {author} {\bibfnamefont {J.}~\bibnamefont {Faraudo}}, \bibinfo {author} {\bibfnamefont {O.}~\bibnamefont {Gang}}, \bibinfo {author} {\bibfnamefont {N.}~\bibnamefont {Gaston}}, \bibinfo {author} {\bibfnamefont {R.}~\bibnamefont {Golestanian}}, \bibinfo {author} {\bibfnamefont {G.~I.}\ \bibnamefont {{Guerrero-Garcia}}}, \bibinfo {author} {\bibfnamefont {M.}~\bibnamefont {Gruenwald}}, \bibinfo {author} {\bibfnamefont {A.}~\bibnamefont {{Haji-Akbari}}}, \bibinfo {author} {\bibfnamefont {M.}~\bibnamefont {Ib{\'a}{\~n}ez}}, \bibinfo {author} {\bibfnamefont {M.}~\bibnamefont {Karg}}, \bibinfo {author} {\bibfnamefont {T.}~\bibnamefont {Kraus}}, \bibinfo {author} {\bibfnamefont {B.}~\bibnamefont {Lee}}, \bibinfo {author} {\bibfnamefont {R.~C.}\ \bibnamefont {Van~Lehn}}, \bibinfo {author} {\bibfnamefont {R.~J.}\ \bibnamefont {Macfarlane}}, \bibinfo {author} {\bibfnamefont {B.~M.}\ \bibnamefont {Mognetti}}, \bibinfo {author} {\bibfnamefont {A.}~\bibnamefont {Nikoubashman}}, \bibinfo {author} {\bibfnamefont {S.}~\bibnamefont {Osat}}, \bibinfo {author} {\bibfnamefont {O.~V.}\ \bibnamefont {Prezhdo}}, \bibinfo {author} {\bibfnamefont {G.~M.}\ \bibnamefont {Rotskoff}}, \bibinfo {author} {\bibfnamefont {L.}~\bibnamefont {Saiz}}, \bibinfo {author} {\bibfnamefont {A.-C.}\ \bibnamefont {Shi}}, \bibinfo {author} {\bibfnamefont {S.}~\bibnamefont {Skrabalak}}, \bibinfo {author} {\bibfnamefont {I.~I.}\ \bibnamefont {Smalyukh}}, \bibinfo {author} {\bibfnamefont {M.}~\bibnamefont {Tagliazucchi}}, \bibinfo {author} {\bibfnamefont {D.~V.}\ \bibnamefont {Talapin}}, \bibinfo {author} {\bibfnamefont {A.~V.}\ \bibnamefont {Tkachenko}}, \bibinfo {author} {\bibfnamefont {S.}~\bibnamefont {Tretiak}}, \bibinfo {author} {\bibfnamefont {D.}~\bibnamefont {Vaknin}}, \bibinfo {author} {\bibfnamefont {A.}~\bibnamefont {{Widmer-Cooper}}}, \bibinfo {author} {\bibfnamefont {G.~C.~L.}\ \bibnamefont {Wong}}, \bibinfo {author} {\bibfnamefont {X.}~\bibnamefont {Ye}}, \bibinfo {author} {\bibfnamefont {S.}~\bibnamefont {Zhou}}, \bibinfo {author} {\bibfnamefont {E.}~\bibnamefont {Rabani}}, \bibinfo {author} {\bibfnamefont {M.}~\bibnamefont {Engel}},\ and\ \bibinfo {author} {\bibfnamefont {A.}~\bibnamefont {Travesset}},\ }\bibfield  {title} {\bibinfo {title} {Nanocrystal {{Assemblies}}: {{Current Advances}} and {{Open Problems}}},\ }\href {https://doi.org/10.1021/acsnano.3c10201} {\bibfield  {journal} {\bibinfo  {journal} {ACS Nano}\ }\textbf {\bibinfo {volume} {18}},\ \bibinfo {pages} {14791} (\bibinfo {year} {2024})}\BibitemShut {NoStop}%
\bibitem [{\citenamefont {Damasceno}\ \emph {et~al.}(2012)\citenamefont {Damasceno}, \citenamefont {Engel},\ and\ \citenamefont {Glotzer}}]{damasceno12}%
  \BibitemOpen
  \bibfield  {author} {\bibinfo {author} {\bibfnamefont {P.~F.}\ \bibnamefont {Damasceno}}, \bibinfo {author} {\bibfnamefont {M.}~\bibnamefont {Engel}},\ and\ \bibinfo {author} {\bibfnamefont {S.~C.}\ \bibnamefont {Glotzer}},\ }\bibfield  {title} {\bibinfo {title} {Predictive {{Self-Assembly}} of {{Polyhedra}} into {{Complex Structures}}},\ }\href {https://doi.org/10.1126/science.1220869} {\bibfield  {journal} {\bibinfo  {journal} {Science}\ }\textbf {\bibinfo {volume} {337}},\ \bibinfo {pages} {453} (\bibinfo {year} {2012})}\BibitemShut {NoStop}%
\bibitem [{\citenamefont {Keys}\ \emph {et~al.}(2011)\citenamefont {Keys}, \citenamefont {Iacovella},\ and\ \citenamefont {Glotzer}}]{keys11}%
  \BibitemOpen
  \bibfield  {author} {\bibinfo {author} {\bibfnamefont {A.~S.}\ \bibnamefont {Keys}}, \bibinfo {author} {\bibfnamefont {C.~R.}\ \bibnamefont {Iacovella}},\ and\ \bibinfo {author} {\bibfnamefont {S.~C.}\ \bibnamefont {Glotzer}},\ }\bibfield  {title} {\bibinfo {title} {Characterizing complex particle morphologies through shape matching: {{Descriptors}}, applications, and algorithms},\ }\href {https://doi.org/10.1016/j.jcp.2011.04.017} {\bibfield  {journal} {\bibinfo  {journal} {J. Comput. Phys.}\ }\textbf {\bibinfo {volume} {230}},\ \bibinfo {pages} {6438} (\bibinfo {year} {2011})}\BibitemShut {NoStop}%
\bibitem [{\citenamefont {Wittkowski}\ and\ \citenamefont {L{\"o}wen}(2011)}]{wittkowski11}%
  \BibitemOpen
  \bibfield  {author} {\bibinfo {author} {\bibfnamefont {R.}~\bibnamefont {Wittkowski}}\ and\ \bibinfo {author} {\bibfnamefont {H.}~\bibnamefont {L{\"o}wen}},\ }\bibfield  {title} {\bibinfo {title} {Dynamical density functional theory for colloidal particles with arbitrary shape},\ }\href {https://doi.org/10.1080/00268976.2011.609145} {\bibfield  {journal} {\bibinfo  {journal} {Mol. Phys.}\ }\textbf {\bibinfo {volume} {109}},\ \bibinfo {pages} {2935} (\bibinfo {year} {2011})}\BibitemShut {NoStop}%
\bibitem [{\citenamefont {Vo}\ and\ \citenamefont {Glotzer}(2022)}]{vo22}%
  \BibitemOpen
  \bibfield  {author} {\bibinfo {author} {\bibfnamefont {T.}~\bibnamefont {Vo}}\ and\ \bibinfo {author} {\bibfnamefont {S.~C.}\ \bibnamefont {Glotzer}},\ }\bibfield  {title} {\bibinfo {title} {A theory of entropic bonding},\ }\href {https://doi.org/10.1073/pnas.2116414119} {\bibfield  {journal} {\bibinfo  {journal} {Proc. Natl. Acad. Sci. U.S.A.}\ }\textbf {\bibinfo {volume} {119}},\ \bibinfo {pages} {e2116414119} (\bibinfo {year} {2022})}\BibitemShut {NoStop}%
\bibitem [{\citenamefont {Anderson}\ \emph {et~al.}(2017)\citenamefont {Anderson}, \citenamefont {Antonaglia}, \citenamefont {Millan}, \citenamefont {Engel},\ and\ \citenamefont {Glotzer}}]{anderson17}%
  \BibitemOpen
  \bibfield  {author} {\bibinfo {author} {\bibfnamefont {J.~A.}\ \bibnamefont {Anderson}}, \bibinfo {author} {\bibfnamefont {J.}~\bibnamefont {Antonaglia}}, \bibinfo {author} {\bibfnamefont {J.~A.}\ \bibnamefont {Millan}}, \bibinfo {author} {\bibfnamefont {M.}~\bibnamefont {Engel}},\ and\ \bibinfo {author} {\bibfnamefont {S.~C.}\ \bibnamefont {Glotzer}},\ }\bibfield  {title} {\bibinfo {title} {Shape and {{Symmetry Determine Two-Dimensional Melting Transitions}} of {{Hard Regular Polygons}}},\ }\href {https://doi.org/10.1103/PhysRevX.7.021001} {\bibfield  {journal} {\bibinfo  {journal} {Phys. Rev. X}\ }\textbf {\bibinfo {volume} {7}},\ \bibinfo {pages} {021001} (\bibinfo {year} {2017})}\BibitemShut {NoStop}%
\bibitem [{\citenamefont {Mizani}\ \emph {et~al.}(2025)\citenamefont {Mizani}, \citenamefont {Oettel}, \citenamefont {Gurin},\ and\ \citenamefont {Varga}}]{mizani25}%
  \BibitemOpen
  \bibfield  {author} {\bibinfo {author} {\bibfnamefont {S.}~\bibnamefont {Mizani}}, \bibinfo {author} {\bibfnamefont {M.}~\bibnamefont {Oettel}}, \bibinfo {author} {\bibfnamefont {P.}~\bibnamefont {Gurin}},\ and\ \bibinfo {author} {\bibfnamefont {S.}~\bibnamefont {Varga}},\ }\bibfield  {title} {\bibinfo {title} {Competition between shape anisotropy and deformation in the ordering and close packing properties of quasi-one-dimensional hard superellipse fluids},\ }\href {https://doi.org/10.1103/jszf-cvdn} {\bibfield  {journal} {\bibinfo  {journal} {Phys. Rev. E}\ }\textbf {\bibinfo {volume} {111}},\ \bibinfo {pages} {064121} (\bibinfo {year} {2025})}\BibitemShut {NoStop}%
\bibitem [{\citenamefont {Bernard}\ and\ \citenamefont {Krauth}(2011)}]{bernard11}%
  \BibitemOpen
  \bibfield  {author} {\bibinfo {author} {\bibfnamefont {E.~P.}\ \bibnamefont {Bernard}}\ and\ \bibinfo {author} {\bibfnamefont {W.}~\bibnamefont {Krauth}},\ }\bibfield  {title} {\bibinfo {title} {Two-{{Step Melting}} in {{Two Dimensions}}: {{First-Order Liquid-Hexatic Transition}}},\ }\href {https://doi.org/10.1103/PhysRevLett.107.155704} {\bibfield  {journal} {\bibinfo  {journal} {Phys. Rev. Lett.}\ }\textbf {\bibinfo {volume} {107}},\ \bibinfo {pages} {155704} (\bibinfo {year} {2011})}\BibitemShut {NoStop}%
\bibitem [{\citenamefont {Engel}\ \emph {et~al.}(2013)\citenamefont {Engel}, \citenamefont {Anderson}, \citenamefont {Glotzer}, \citenamefont {Isobe}, \citenamefont {Bernard},\ and\ \citenamefont {Krauth}}]{engel13}%
  \BibitemOpen
  \bibfield  {author} {\bibinfo {author} {\bibfnamefont {M.}~\bibnamefont {Engel}}, \bibinfo {author} {\bibfnamefont {J.~A.}\ \bibnamefont {Anderson}}, \bibinfo {author} {\bibfnamefont {S.~C.}\ \bibnamefont {Glotzer}}, \bibinfo {author} {\bibfnamefont {M.}~\bibnamefont {Isobe}}, \bibinfo {author} {\bibfnamefont {E.~P.}\ \bibnamefont {Bernard}},\ and\ \bibinfo {author} {\bibfnamefont {W.}~\bibnamefont {Krauth}},\ }\bibfield  {title} {\bibinfo {title} {Hard-disk equation of state: {{First-order}} liquid-hexatic transition in two dimensions with three simulation methods},\ }\href {https://doi.org/10.1103/PhysRevE.87.042134} {\bibfield  {journal} {\bibinfo  {journal} {Phys. Rev. E}\ }\textbf {\bibinfo {volume} {87}},\ \bibinfo {pages} {042134} (\bibinfo {year} {2013})}\BibitemShut {NoStop}%
\bibitem [{\citenamefont {Kapfer}\ and\ \citenamefont {Krauth}(2015)}]{kapfer15}%
  \BibitemOpen
  \bibfield  {author} {\bibinfo {author} {\bibfnamefont {S.~C.}\ \bibnamefont {Kapfer}}\ and\ \bibinfo {author} {\bibfnamefont {W.}~\bibnamefont {Krauth}},\ }\bibfield  {title} {\bibinfo {title} {Two-{{Dimensional Melting}}: {{From Liquid-Hexatic Coexistence}} to {{Continuous Transitions}}},\ }\href {https://doi.org/10.1103/PhysRevLett.114.035702} {\bibfield  {journal} {\bibinfo  {journal} {Phys. Rev. Lett.}\ }\textbf {\bibinfo {volume} {114}},\ \bibinfo {pages} {035702} (\bibinfo {year} {2015})}\BibitemShut {NoStop}%
\bibitem [{\citenamefont {Dertli}\ and\ \citenamefont {Speck}(2025)}]{dertli25}%
  \BibitemOpen
  \bibfield  {author} {\bibinfo {author} {\bibfnamefont {D.}~\bibnamefont {Dertli}}\ and\ \bibinfo {author} {\bibfnamefont {T.}~\bibnamefont {Speck}},\ }\bibfield  {title} {\bibinfo {title} {In pursuit of the tetratic phase in hard rectangles},\ }\href {https://doi.org/10.1103/PhysRevResearch.7.L012034} {\bibfield  {journal} {\bibinfo  {journal} {Phys. Rev. Research}\ }\textbf {\bibinfo {volume} {7}},\ \bibinfo {pages} {L012034} (\bibinfo {year} {2025})}\BibitemShut {NoStop}%
\bibitem [{\citenamefont {Sun}\ \emph {et~al.}(2012)\citenamefont {Sun}, \citenamefont {Evans}, \citenamefont {Lee}, \citenamefont {Senyuk}, \citenamefont {Keller}, \citenamefont {He},\ and\ \citenamefont {Smalyukh}}]{sun12}%
  \BibitemOpen
  \bibfield  {author} {\bibinfo {author} {\bibfnamefont {Y.}~\bibnamefont {Sun}}, \bibinfo {author} {\bibfnamefont {J.~S.}\ \bibnamefont {Evans}}, \bibinfo {author} {\bibfnamefont {T.}~\bibnamefont {Lee}}, \bibinfo {author} {\bibfnamefont {B.}~\bibnamefont {Senyuk}}, \bibinfo {author} {\bibfnamefont {P.}~\bibnamefont {Keller}}, \bibinfo {author} {\bibfnamefont {S.}~\bibnamefont {He}},\ and\ \bibinfo {author} {\bibfnamefont {I.~I.}\ \bibnamefont {Smalyukh}},\ }\bibfield  {title} {\bibinfo {title} {Optical manipulation of shape-morphing elastomeric liquid crystal microparticles doped with gold nanocrystals},\ }\bibfield  {journal} {\bibinfo  {journal} {Appl. Phys. Lett.}\ }\textbf {\bibinfo {volume} {100}},\ \href {https://doi.org/10.1063/1.4729143} {10.1063/1.4729143} (\bibinfo {year} {2012})\BibitemShut {NoStop}%
\bibitem [{\citenamefont {Epstein}\ \emph {et~al.}(2015)\citenamefont {Epstein}, \citenamefont {Yoon}, \citenamefont {Madhukar}, \citenamefont {Hsia},\ and\ \citenamefont {Braun}}]{epstein15}%
  \BibitemOpen
  \bibfield  {author} {\bibinfo {author} {\bibfnamefont {E.}~\bibnamefont {Epstein}}, \bibinfo {author} {\bibfnamefont {J.}~\bibnamefont {Yoon}}, \bibinfo {author} {\bibfnamefont {A.}~\bibnamefont {Madhukar}}, \bibinfo {author} {\bibfnamefont {K.~J.}\ \bibnamefont {Hsia}},\ and\ \bibinfo {author} {\bibfnamefont {P.~V.}\ \bibnamefont {Braun}},\ }\bibfield  {title} {\bibinfo {title} {Colloidal {{Particles}} that {{Rapidly Change Shape}} via {{Elastic Instabilities}}},\ }\href {https://doi.org/10.1002/smll.201502198} {\bibfield  {journal} {\bibinfo  {journal} {Small}\ }\textbf {\bibinfo {volume} {11}},\ \bibinfo {pages} {6051} (\bibinfo {year} {2015})}\BibitemShut {NoStop}%
\bibitem [{\citenamefont {Tanjeem}\ \emph {et~al.}(2022)\citenamefont {Tanjeem}, \citenamefont {Minnis}, \citenamefont {Hayward},\ and\ \citenamefont {Shields}}]{tanjeem22}%
  \BibitemOpen
  \bibfield  {author} {\bibinfo {author} {\bibfnamefont {N.}~\bibnamefont {Tanjeem}}, \bibinfo {author} {\bibfnamefont {M.~B.}\ \bibnamefont {Minnis}}, \bibinfo {author} {\bibfnamefont {R.~C.}\ \bibnamefont {Hayward}},\ and\ \bibinfo {author} {\bibfnamefont {C.~W.}\ \bibnamefont {Shields}},\ }\bibfield  {title} {\bibinfo {title} {Shape-{{Changing Particles}}: {{From Materials Design}} and {{Mechanisms}} to {{Implementation}}},\ }\bibfield  {journal} {\bibinfo  {journal} {Adv. Mater.}\ }\textbf {\bibinfo {volume} {34}},\ \href {https://doi.org/10.1002/adma.202105758} {10.1002/adma.202105758} (\bibinfo {year} {2022})\BibitemShut {NoStop}%
\bibitem [{\citenamefont {Zhan}\ \emph {et~al.}(2023)\citenamefont {Zhan}, \citenamefont {Peil}, \citenamefont {Jiang}, \citenamefont {Wang}, \citenamefont {Mousavi}, \citenamefont {Xiong}, \citenamefont {Shen}, \citenamefont {Shang}, \citenamefont {Ding}, \citenamefont {Lin}, \citenamefont {Ke},\ and\ \citenamefont {Liu}}]{zhan23}%
  \BibitemOpen
  \bibfield  {author} {\bibinfo {author} {\bibfnamefont {P.}~\bibnamefont {Zhan}}, \bibinfo {author} {\bibfnamefont {A.}~\bibnamefont {Peil}}, \bibinfo {author} {\bibfnamefont {Q.}~\bibnamefont {Jiang}}, \bibinfo {author} {\bibfnamefont {D.}~\bibnamefont {Wang}}, \bibinfo {author} {\bibfnamefont {S.}~\bibnamefont {Mousavi}}, \bibinfo {author} {\bibfnamefont {Q.}~\bibnamefont {Xiong}}, \bibinfo {author} {\bibfnamefont {Q.}~\bibnamefont {Shen}}, \bibinfo {author} {\bibfnamefont {Y.}~\bibnamefont {Shang}}, \bibinfo {author} {\bibfnamefont {B.}~\bibnamefont {Ding}}, \bibinfo {author} {\bibfnamefont {C.}~\bibnamefont {Lin}}, \bibinfo {author} {\bibfnamefont {Y.}~\bibnamefont {Ke}},\ and\ \bibinfo {author} {\bibfnamefont {N.}~\bibnamefont {Liu}},\ }\bibfield  {title} {\bibinfo {title} {Recent {{Advances}} in {{DNA Origami-Engineered Nanomaterials}} and {{Applications}}},\ }\href {https://doi.org/10.1021/acs.chemrev.3c00028} {\bibfield  {journal} {\bibinfo  {journal} {Chem. Rev.}\ }\textbf {\bibinfo {volume} {123}},\ \bibinfo {pages} {3976} (\bibinfo {year} {2023})}\BibitemShut {NoStop}%
\bibitem [{\citenamefont {Rothemund}(2006)}]{rothemund06}%
  \BibitemOpen
  \bibfield  {author} {\bibinfo {author} {\bibfnamefont {P.~W.~K.}\ \bibnamefont {Rothemund}},\ }\bibfield  {title} {\bibinfo {title} {Folding {{DNA}} to create nanoscale shapes and patterns},\ }\href {https://doi.org/10.1038/nature04586} {\bibfield  {journal} {\bibinfo  {journal} {Nature}\ }\textbf {\bibinfo {volume} {440}},\ \bibinfo {pages} {297} (\bibinfo {year} {2006})}\BibitemShut {NoStop}%
\bibitem [{\citenamefont {Dietz}\ \emph {et~al.}(2009)\citenamefont {Dietz}, \citenamefont {Douglas},\ and\ \citenamefont {Shih}}]{dietz09}%
  \BibitemOpen
  \bibfield  {author} {\bibinfo {author} {\bibfnamefont {H.}~\bibnamefont {Dietz}}, \bibinfo {author} {\bibfnamefont {S.~M.}\ \bibnamefont {Douglas}},\ and\ \bibinfo {author} {\bibfnamefont {W.~M.}\ \bibnamefont {Shih}},\ }\bibfield  {title} {\bibinfo {title} {Folding {{DNA}} into {{Twisted}} and {{Curved Nanoscale Shapes}}},\ }\href {https://doi.org/10.1126/science.1174251} {\bibfield  {journal} {\bibinfo  {journal} {Science}\ }\textbf {\bibinfo {volume} {325}},\ \bibinfo {pages} {725} (\bibinfo {year} {2009})}\BibitemShut {NoStop}%
\bibitem [{\citenamefont {Hong}\ \emph {et~al.}(2017)\citenamefont {Hong}, \citenamefont {Zhang}, \citenamefont {Liu},\ and\ \citenamefont {Yan}}]{hong17}%
  \BibitemOpen
  \bibfield  {author} {\bibinfo {author} {\bibfnamefont {F.}~\bibnamefont {Hong}}, \bibinfo {author} {\bibfnamefont {F.}~\bibnamefont {Zhang}}, \bibinfo {author} {\bibfnamefont {Y.}~\bibnamefont {Liu}},\ and\ \bibinfo {author} {\bibfnamefont {H.}~\bibnamefont {Yan}},\ }\bibfield  {title} {\bibinfo {title} {{{DNA Origami}}: {{Scaffolds}} for {{Creating Higher Order Structures}}},\ }\href {https://doi.org/10.1021/acs.chemrev.6b00825} {\bibfield  {journal} {\bibinfo  {journal} {Chem. Rev.}\ }\textbf {\bibinfo {volume} {117}},\ \bibinfo {pages} {12584} (\bibinfo {year} {2017})}\BibitemShut {NoStop}%
\bibitem [{\citenamefont {Endo}\ \emph {et~al.}(2013)\citenamefont {Endo}, \citenamefont {Yang},\ and\ \citenamefont {Sugiyama}}]{endo13}%
  \BibitemOpen
  \bibfield  {author} {\bibinfo {author} {\bibfnamefont {M.}~\bibnamefont {Endo}}, \bibinfo {author} {\bibfnamefont {Y.}~\bibnamefont {Yang}},\ and\ \bibinfo {author} {\bibfnamefont {H.}~\bibnamefont {Sugiyama}},\ }\bibfield  {title} {\bibinfo {title} {{{DNA}} origami technology for biomaterials applications},\ }\href {https://doi.org/10.1039/C2BM00154C} {\bibfield  {journal} {\bibinfo  {journal} {Biomater. Sci.}\ }\textbf {\bibinfo {volume} {1}},\ \bibinfo {pages} {347} (\bibinfo {year} {2013})}\BibitemShut {NoStop}%
\bibitem [{\citenamefont {Sun}\ \emph {et~al.}(2022)\citenamefont {Sun}, \citenamefont {Sun}, \citenamefont {Xiao}, \citenamefont {Lai}, \citenamefont {Li}, \citenamefont {Fan},\ and\ \citenamefont {Pei}}]{sun22}%
  \BibitemOpen
  \bibfield  {author} {\bibinfo {author} {\bibfnamefont {Y.}~\bibnamefont {Sun}}, \bibinfo {author} {\bibfnamefont {J.}~\bibnamefont {Sun}}, \bibinfo {author} {\bibfnamefont {M.}~\bibnamefont {Xiao}}, \bibinfo {author} {\bibfnamefont {W.}~\bibnamefont {Lai}}, \bibinfo {author} {\bibfnamefont {L.}~\bibnamefont {Li}}, \bibinfo {author} {\bibfnamefont {C.}~\bibnamefont {Fan}},\ and\ \bibinfo {author} {\bibfnamefont {H.}~\bibnamefont {Pei}},\ }\bibfield  {title} {\bibinfo {title} {{{DNA}} origami--based artificial antigen-presenting cells for adoptive {{T}} cell therapy},\ }\href {https://doi.org/10.1126/sciadv.add1106} {\bibfield  {journal} {\bibinfo  {journal} {Sci. Adv.}\ }\textbf {\bibinfo {volume} {8}},\ \bibinfo {pages} {eadd1106} (\bibinfo {year} {2022})}\BibitemShut {NoStop}%
\bibitem [{\citenamefont {He}\ \emph {et~al.}(2023)\citenamefont {He}, \citenamefont {Fan}, \citenamefont {Wang}, \citenamefont {Wei},\ and\ \citenamefont {Hu}}]{he23}%
  \BibitemOpen
  \bibfield  {author} {\bibinfo {author} {\bibfnamefont {S.}~\bibnamefont {He}}, \bibinfo {author} {\bibfnamefont {T.}~\bibnamefont {Fan}}, \bibinfo {author} {\bibfnamefont {Y.}~\bibnamefont {Wang}}, \bibinfo {author} {\bibfnamefont {C.}~\bibnamefont {Wei}},\ and\ \bibinfo {author} {\bibfnamefont {J.}~\bibnamefont {Hu}},\ }\bibfield  {title} {\bibinfo {title} {Recent {{Advances}} in {{DNA Nanostructure}}-enabled {{Drug Delivery}}},\ }\href {https://doi.org/10.1002/cnma.202200459} {\bibfield  {journal} {\bibinfo  {journal} {ChemNanoMat}\ }\textbf {\bibinfo {volume} {9}},\ \bibinfo {pages} {e202200459} (\bibinfo {year} {2023})}\BibitemShut {NoStop}%
\bibitem [{\citenamefont {Zhang}\ \emph {et~al.}(2023)\citenamefont {Zhang}, \citenamefont {Tian}, \citenamefont {Wang}, \citenamefont {Wang}, \citenamefont {Liu}, \citenamefont {Long},\ and\ \citenamefont {Jiang}}]{zhang23}%
  \BibitemOpen
  \bibfield  {author} {\bibinfo {author} {\bibfnamefont {Y.}~\bibnamefont {Zhang}}, \bibinfo {author} {\bibfnamefont {X.}~\bibnamefont {Tian}}, \bibinfo {author} {\bibfnamefont {Z.}~\bibnamefont {Wang}}, \bibinfo {author} {\bibfnamefont {H.}~\bibnamefont {Wang}}, \bibinfo {author} {\bibfnamefont {F.}~\bibnamefont {Liu}}, \bibinfo {author} {\bibfnamefont {Q.}~\bibnamefont {Long}},\ and\ \bibinfo {author} {\bibfnamefont {S.}~\bibnamefont {Jiang}},\ }\bibfield  {title} {\bibinfo {title} {Advanced applications of {{DNA}} nanostructures dominated by {{DNA}} origami in antitumor drug delivery},\ }\href {https://doi.org/10.3389/fmolb.2023.1239952} {\bibfield  {journal} {\bibinfo  {journal} {Front. Mol. Biosci.}\ }\textbf {\bibinfo {volume} {10}},\ \bibinfo {pages} {1239952} (\bibinfo {year} {2023})}\BibitemShut {NoStop}%
\bibitem [{\citenamefont {Safarkhani}\ \emph {et~al.}(2024)\citenamefont {Safarkhani}, \citenamefont {Ahmadi}, \citenamefont {Ipakchi}, \citenamefont {Saeb}, \citenamefont {Makvandi}, \citenamefont {Ebrahimi~Warkiani}, \citenamefont {Rabiee},\ and\ \citenamefont {Huh}}]{safarkhani24}%
  \BibitemOpen
  \bibfield  {author} {\bibinfo {author} {\bibfnamefont {M.}~\bibnamefont {Safarkhani}}, \bibinfo {author} {\bibfnamefont {S.}~\bibnamefont {Ahmadi}}, \bibinfo {author} {\bibfnamefont {H.}~\bibnamefont {Ipakchi}}, \bibinfo {author} {\bibfnamefont {M.~R.}\ \bibnamefont {Saeb}}, \bibinfo {author} {\bibfnamefont {P.}~\bibnamefont {Makvandi}}, \bibinfo {author} {\bibfnamefont {M.}~\bibnamefont {Ebrahimi~Warkiani}}, \bibinfo {author} {\bibfnamefont {N.}~\bibnamefont {Rabiee}},\ and\ \bibinfo {author} {\bibfnamefont {Y.}~\bibnamefont {Huh}},\ }\bibfield  {title} {\bibinfo {title} {Advancements in {{Aptamer}}-{{Driven DNA Nanostructures}} for {{Precision Drug Delivery}}},\ }\href {https://doi.org/10.1002/advs.202401617} {\bibfield  {journal} {\bibinfo  {journal} {Advanced Science}\ }\textbf {\bibinfo {volume} {11}},\ \bibinfo {pages} {2401617} (\bibinfo {year} {2024})}\BibitemShut {NoStop}%
\bibitem [{\citenamefont {Ding}\ \emph {et~al.}(2025)\citenamefont {Ding}, \citenamefont {Liu}, \citenamefont {Peil}, \citenamefont {Fan}, \citenamefont {Chao},\ and\ \citenamefont {Liu}}]{ding25}%
  \BibitemOpen
  \bibfield  {author} {\bibinfo {author} {\bibfnamefont {L.}~\bibnamefont {Ding}}, \bibinfo {author} {\bibfnamefont {B.}~\bibnamefont {Liu}}, \bibinfo {author} {\bibfnamefont {A.}~\bibnamefont {Peil}}, \bibinfo {author} {\bibfnamefont {S.}~\bibnamefont {Fan}}, \bibinfo {author} {\bibfnamefont {J.}~\bibnamefont {Chao}},\ and\ \bibinfo {author} {\bibfnamefont {N.}~\bibnamefont {Liu}},\ }\bibfield  {title} {\bibinfo {title} {{{DNA}}-{{Directed Assembly}} of {{Photonic Nanomaterials}} for {{Diagnostic}} and {{Therapeutic Applications}}},\ }\href {https://doi.org/10.1002/adma.202500086} {\bibfield  {journal} {\bibinfo  {journal} {Advanced Materials}\ ,\ \bibinfo {pages} {2500086}} (\bibinfo {year} {2025})}\BibitemShut {NoStop}%
\bibitem [{\citenamefont {Khmelinskaia}\ \emph {et~al.}(2021)\citenamefont {Khmelinskaia}, \citenamefont {Franquelim}, \citenamefont {Yaadav}, \citenamefont {Petrov},\ and\ \citenamefont {Schwille}}]{khmelinskaia21}%
  \BibitemOpen
  \bibfield  {author} {\bibinfo {author} {\bibfnamefont {A.}~\bibnamefont {Khmelinskaia}}, \bibinfo {author} {\bibfnamefont {H.~G.}\ \bibnamefont {Franquelim}}, \bibinfo {author} {\bibfnamefont {R.}~\bibnamefont {Yaadav}}, \bibinfo {author} {\bibfnamefont {E.~P.}\ \bibnamefont {Petrov}},\ and\ \bibinfo {author} {\bibfnamefont {P.}~\bibnamefont {Schwille}},\ }\bibfield  {title} {\bibinfo {title} {Membrane-{{Mediated Self-Organization}} of {{Rod-Like DNA Origami}} on {{Supported Lipid Bilayers}}},\ }\href {https://doi.org/10.1002/admi.202101094} {\bibfield  {journal} {\bibinfo  {journal} {Adv. Mater. Interfaces}\ }\textbf {\bibinfo {volume} {8}},\ \bibinfo {pages} {2101094} (\bibinfo {year} {2021})}\BibitemShut {NoStop}%
\bibitem [{\citenamefont {Fan}\ \emph {et~al.}(2025)\citenamefont {Fan}, \citenamefont {Wang}, \citenamefont {Ding}, \citenamefont {Speck}, \citenamefont {Yan}, \citenamefont {Nussberger},\ and\ \citenamefont {Liu}}]{fan25}%
  \BibitemOpen
  \bibfield  {author} {\bibinfo {author} {\bibfnamefont {S.}~\bibnamefont {Fan}}, \bibinfo {author} {\bibfnamefont {S.}~\bibnamefont {Wang}}, \bibinfo {author} {\bibfnamefont {L.}~\bibnamefont {Ding}}, \bibinfo {author} {\bibfnamefont {T.}~\bibnamefont {Speck}}, \bibinfo {author} {\bibfnamefont {H.}~\bibnamefont {Yan}}, \bibinfo {author} {\bibfnamefont {S.}~\bibnamefont {Nussberger}},\ and\ \bibinfo {author} {\bibfnamefont {N.}~\bibnamefont {Liu}},\ }\bibfield  {title} {\bibinfo {title} {Morphology remodelling and membrane channel formation in synthetic cells via reconfigurable {{DNA}} nanorafts},\ }\href {https://doi.org/10.1038/s41563-024-02075-9} {\bibfield  {journal} {\bibinfo  {journal} {Nat. Mater.}\ }\textbf {\bibinfo {volume} {24}},\ \bibinfo {pages} {278} (\bibinfo {year} {2025})}\BibitemShut {NoStop}%
\bibitem [{\citenamefont {Zhang}\ and\ \citenamefont {Seelig}(2011)}]{zhang11}%
  \BibitemOpen
  \bibfield  {author} {\bibinfo {author} {\bibfnamefont {D.~Y.}\ \bibnamefont {Zhang}}\ and\ \bibinfo {author} {\bibfnamefont {G.}~\bibnamefont {Seelig}},\ }\bibfield  {title} {\bibinfo {title} {Dynamic {{DNA}} nanotechnology using strand-displacement reactions},\ }\href {https://doi.org/10.1038/nchem.957} {\bibfield  {journal} {\bibinfo  {journal} {Nature Chem}\ }\textbf {\bibinfo {volume} {3}},\ \bibinfo {pages} {103} (\bibinfo {year} {2011})}\BibitemShut {NoStop}%
\bibitem [{\citenamefont {Song}\ \emph {et~al.}(2017)\citenamefont {Song}, \citenamefont {Li}, \citenamefont {Wang}, \citenamefont {Meyer}, \citenamefont {Mao},\ and\ \citenamefont {Ke}}]{song17}%
  \BibitemOpen
  \bibfield  {author} {\bibinfo {author} {\bibfnamefont {J.}~\bibnamefont {Song}}, \bibinfo {author} {\bibfnamefont {Z.}~\bibnamefont {Li}}, \bibinfo {author} {\bibfnamefont {P.}~\bibnamefont {Wang}}, \bibinfo {author} {\bibfnamefont {T.}~\bibnamefont {Meyer}}, \bibinfo {author} {\bibfnamefont {C.}~\bibnamefont {Mao}},\ and\ \bibinfo {author} {\bibfnamefont {Y.}~\bibnamefont {Ke}},\ }\bibfield  {title} {\bibinfo {title} {Reconfiguration of {{DNA}} molecular arrays driven by information relay},\ }\href {https://doi.org/10.1126/science.aan3377} {\bibfield  {journal} {\bibinfo  {journal} {Science}\ }\textbf {\bibinfo {volume} {357}},\ \bibinfo {pages} {eaan3377} (\bibinfo {year} {2017})}\BibitemShut {NoStop}%
\bibitem [{\citenamefont {Wang}\ \emph {et~al.}(2020)\citenamefont {Wang}, \citenamefont {Yu}, \citenamefont {Ji}, \citenamefont {Chang}, \citenamefont {Song},\ and\ \citenamefont {Ke}}]{wang20}%
  \BibitemOpen
  \bibfield  {author} {\bibinfo {author} {\bibfnamefont {D.}~\bibnamefont {Wang}}, \bibinfo {author} {\bibfnamefont {L.}~\bibnamefont {Yu}}, \bibinfo {author} {\bibfnamefont {B.}~\bibnamefont {Ji}}, \bibinfo {author} {\bibfnamefont {S.}~\bibnamefont {Chang}}, \bibinfo {author} {\bibfnamefont {J.}~\bibnamefont {Song}},\ and\ \bibinfo {author} {\bibfnamefont {Y.}~\bibnamefont {Ke}},\ }\bibfield  {title} {\bibinfo {title} {Programming the {{Curvatures}} in {{Reconfigurable DNA Domino Origami}} by {{Using Asymmetric Units}}},\ }\href {https://doi.org/10.1021/acs.nanolett.0c03348} {\bibfield  {journal} {\bibinfo  {journal} {Nano Lett.}\ }\textbf {\bibinfo {volume} {20}},\ \bibinfo {pages} {8236} (\bibinfo {year} {2020})}\BibitemShut {NoStop}%
\bibitem [{\citenamefont {Kim}\ \emph {et~al.}(2023)\citenamefont {Kim}, \citenamefont {Lee}, \citenamefont {Jeon}, \citenamefont {Lee}, \citenamefont {Kim}, \citenamefont {Lee}, \citenamefont {Kim}, \citenamefont {Cho},\ and\ \citenamefont {Kim}}]{kim23}%
  \BibitemOpen
  \bibfield  {author} {\bibinfo {author} {\bibfnamefont {M.}~\bibnamefont {Kim}}, \bibinfo {author} {\bibfnamefont {C.}~\bibnamefont {Lee}}, \bibinfo {author} {\bibfnamefont {K.}~\bibnamefont {Jeon}}, \bibinfo {author} {\bibfnamefont {J.~Y.}\ \bibnamefont {Lee}}, \bibinfo {author} {\bibfnamefont {Y.-J.}\ \bibnamefont {Kim}}, \bibinfo {author} {\bibfnamefont {J.~G.}\ \bibnamefont {Lee}}, \bibinfo {author} {\bibfnamefont {H.}~\bibnamefont {Kim}}, \bibinfo {author} {\bibfnamefont {M.}~\bibnamefont {Cho}},\ and\ \bibinfo {author} {\bibfnamefont {D.-N.}\ \bibnamefont {Kim}},\ }\bibfield  {title} {\bibinfo {title} {Harnessing a paper-folding mechanism for reconfigurable {{DNA}} origami},\ }\href {https://doi.org/10.1038/s41586-023-06181-7} {\bibfield  {journal} {\bibinfo  {journal} {Nature}\ }\textbf {\bibinfo {volume} {619}},\ \bibinfo {pages} {78} (\bibinfo {year} {2023})}\BibitemShut {NoStop}%
\bibitem [{\citenamefont {Frenkel}\ and\ \citenamefont {Smit}(2023)}]{frenkel23}%
  \BibitemOpen
  \bibfield  {author} {\bibinfo {author} {\bibfnamefont {D.}~\bibnamefont {Frenkel}}\ and\ \bibinfo {author} {\bibfnamefont {B.}~\bibnamefont {Smit}},\ }\href@noop {} {\emph {\bibinfo {title} {Understanding Molecular Simulation: From Algorithms to Applications}}}\ (\bibinfo  {publisher} {Elsevier},\ \bibinfo {year} {2023})\BibitemShut {NoStop}%
\bibitem [{\citenamefont {Anderson}\ \emph {et~al.}(2008)\citenamefont {Anderson}, \citenamefont {Lorenz},\ and\ \citenamefont {Travesset}}]{anderson08}%
  \BibitemOpen
  \bibfield  {author} {\bibinfo {author} {\bibfnamefont {J.~A.}\ \bibnamefont {Anderson}}, \bibinfo {author} {\bibfnamefont {C.~D.}\ \bibnamefont {Lorenz}},\ and\ \bibinfo {author} {\bibfnamefont {A.}~\bibnamefont {Travesset}},\ }\bibfield  {title} {\bibinfo {title} {General purpose molecular dynamics simulations fully implemented on graphics processing units},\ }\href {https://doi.org/10.1016/j.jcp.2008.01.047} {\bibfield  {journal} {\bibinfo  {journal} {J. Comput. Phys.}\ }\textbf {\bibinfo {volume} {227}},\ \bibinfo {pages} {5342} (\bibinfo {year} {2008})}\BibitemShut {NoStop}%
\bibitem [{\citenamefont {Glaser}\ \emph {et~al.}(2015)\citenamefont {Glaser}, \citenamefont {Nguyen}, \citenamefont {Anderson}, \citenamefont {Lui}, \citenamefont {Spiga}, \citenamefont {Millan}, \citenamefont {Morse},\ and\ \citenamefont {Glotzer}}]{glaser15}%
  \BibitemOpen
  \bibfield  {author} {\bibinfo {author} {\bibfnamefont {J.}~\bibnamefont {Glaser}}, \bibinfo {author} {\bibfnamefont {T.~D.}\ \bibnamefont {Nguyen}}, \bibinfo {author} {\bibfnamefont {J.~A.}\ \bibnamefont {Anderson}}, \bibinfo {author} {\bibfnamefont {P.}~\bibnamefont {Lui}}, \bibinfo {author} {\bibfnamefont {F.}~\bibnamefont {Spiga}}, \bibinfo {author} {\bibfnamefont {J.~A.}\ \bibnamefont {Millan}}, \bibinfo {author} {\bibfnamefont {D.~C.}\ \bibnamefont {Morse}},\ and\ \bibinfo {author} {\bibfnamefont {S.~C.}\ \bibnamefont {Glotzer}},\ }\bibfield  {title} {\bibinfo {title} {Strong scaling of general-purpose molecular dynamics simulations on {{GPUs}}},\ }\href {https://doi.org/10.1016/j.cpc.2015.02.028} {\bibfield  {journal} {\bibinfo  {journal} {Comput. Phys. Commun.}\ }\textbf {\bibinfo {volume} {192}},\ \bibinfo {pages} {97} (\bibinfo {year} {2015})}\BibitemShut {NoStop}%
\bibitem [{\citenamefont {Anderson}\ \emph {et~al.}(2016)\citenamefont {Anderson}, \citenamefont {Eric~Irrgang},\ and\ \citenamefont {Glotzer}}]{anderson16}%
  \BibitemOpen
  \bibfield  {author} {\bibinfo {author} {\bibfnamefont {J.~A.}\ \bibnamefont {Anderson}}, \bibinfo {author} {\bibfnamefont {M.}~\bibnamefont {Eric~Irrgang}},\ and\ \bibinfo {author} {\bibfnamefont {S.~C.}\ \bibnamefont {Glotzer}},\ }\bibfield  {title} {\bibinfo {title} {Scalable {{Metropolis Monte Carlo}} for simulation of hard shapes},\ }\href {https://doi.org/10.1016/j.cpc.2016.02.024} {\bibfield  {journal} {\bibinfo  {journal} {Comput. Phys. Commun.}\ }\textbf {\bibinfo {volume} {204}},\ \bibinfo {pages} {21} (\bibinfo {year} {2016})}\BibitemShut {NoStop}%
\bibitem [{\citenamefont {Anderson}\ \emph {et~al.}(2020)\citenamefont {Anderson}, \citenamefont {Glaser},\ and\ \citenamefont {Glotzer}}]{anderson20}%
  \BibitemOpen
  \bibfield  {author} {\bibinfo {author} {\bibfnamefont {J.~A.}\ \bibnamefont {Anderson}}, \bibinfo {author} {\bibfnamefont {J.}~\bibnamefont {Glaser}},\ and\ \bibinfo {author} {\bibfnamefont {S.~C.}\ \bibnamefont {Glotzer}},\ }\bibfield  {title} {\bibinfo {title} {{{HOOMD-blue}}: {{A Python}} package for high-performance molecular dynamics and hard particle {{Monte Carlo}} simulations},\ }\href {https://doi.org/10.1016/j.commatsci.2019.109363} {\bibfield  {journal} {\bibinfo  {journal} {Comput. Mater. Sci.}\ }\textbf {\bibinfo {volume} {173}},\ \bibinfo {pages} {109363} (\bibinfo {year} {2020})}\BibitemShut {NoStop}%
\bibitem [{\citenamefont {Jack}\ \emph {et~al.}(2015)\citenamefont {Jack}, \citenamefont {Thompson},\ and\ \citenamefont {Sollich}}]{jack15}%
  \BibitemOpen
  \bibfield  {author} {\bibinfo {author} {\bibfnamefont {R.~L.}\ \bibnamefont {Jack}}, \bibinfo {author} {\bibfnamefont {I.~R.}\ \bibnamefont {Thompson}},\ and\ \bibinfo {author} {\bibfnamefont {P.}~\bibnamefont {Sollich}},\ }\bibfield  {title} {\bibinfo {title} {Hyperuniformity and {{Phase Separation}} in {{Biased Ensembles}} of {{Trajectories}} for {{Diffusive Systems}}},\ }\href {https://doi.org/10.1103/PhysRevLett.114.060601} {\bibfield  {journal} {\bibinfo  {journal} {Phys. Rev. Lett.}\ }\textbf {\bibinfo {volume} {114}},\ \bibinfo {pages} {060601} (\bibinfo {year} {2015})}\BibitemShut {NoStop}%
\bibitem [{sm()}]{sm}%
  \BibitemOpen
  \bibfield  {title} {\bibinfo {title} {Supplemental information},\ }\bibinfo {note} {supplemental material at xxx containing additional information on the simulations}\BibitemShut {NoStop}%
\bibitem [{\citenamefont {Ramasubramani}\ \emph {et~al.}(2020)\citenamefont {Ramasubramani}, \citenamefont {Dice}, \citenamefont {Harper}, \citenamefont {Spellings}, \citenamefont {Anderson},\ and\ \citenamefont {Glotzer}}]{ramasubramani20}%
  \BibitemOpen
  \bibfield  {author} {\bibinfo {author} {\bibfnamefont {V.}~\bibnamefont {Ramasubramani}}, \bibinfo {author} {\bibfnamefont {B.~D.}\ \bibnamefont {Dice}}, \bibinfo {author} {\bibfnamefont {E.~S.}\ \bibnamefont {Harper}}, \bibinfo {author} {\bibfnamefont {M.~P.}\ \bibnamefont {Spellings}}, \bibinfo {author} {\bibfnamefont {J.~A.}\ \bibnamefont {Anderson}},\ and\ \bibinfo {author} {\bibfnamefont {S.~C.}\ \bibnamefont {Glotzer}},\ }\bibfield  {title} {\bibinfo {title} {Freud: {{A}} software suite for high throughput analysis of particle simulation data},\ }\href {https://doi.org/10.1016/j.cpc.2020.107275} {\bibfield  {journal} {\bibinfo  {journal} {Comput. Phys. Commun.}\ }\textbf {\bibinfo {volume} {254}},\ \bibinfo {pages} {107275} (\bibinfo {year} {2020})}\BibitemShut {NoStop}%
\bibitem [{\citenamefont {Lazaro}\ and\ \citenamefont {Hagan}(2016)}]{lazaro16}%
  \BibitemOpen
  \bibfield  {author} {\bibinfo {author} {\bibfnamefont {G.~R.}\ \bibnamefont {Lazaro}}\ and\ \bibinfo {author} {\bibfnamefont {M.~F.}\ \bibnamefont {Hagan}},\ }\bibfield  {title} {\bibinfo {title} {Allosteric {{Control}} of {{Icosahedral Capsid Assembly}}},\ }\href {https://doi.org/10.1021/acs.jpcb.6b02768} {\bibfield  {journal} {\bibinfo  {journal} {J. Phys. Chem. B}\ }\textbf {\bibinfo {volume} {120}},\ \bibinfo {pages} {6306} (\bibinfo {year} {2016})}\BibitemShut {NoStop}%
\bibitem [{\citenamefont {Metson}(2025)}]{metson25}%
  \BibitemOpen
  \bibfield  {author} {\bibinfo {author} {\bibfnamefont {J.}~\bibnamefont {Metson}},\ }\bibfield  {title} {\bibinfo {title} {Designing complex behaviors using transition-based allosteric self-assembly},\ }\href {https://doi.org/10.1103/hx18-tm3p} {\bibfield  {journal} {\bibinfo  {journal} {Phys. Rev. Research}\ }\textbf {\bibinfo {volume} {7}},\ \bibinfo {pages} {033044} (\bibinfo {year} {2025})}\BibitemShut {NoStop}%
\end{thebibliography}
\end{document}